\documentclass[10pt,conference]{IEEEtran}

\usepackage{graphicx}%
\usepackage{multirow}%
\usepackage{amsmath,amssymb,amsfonts}%
\usepackage{mathrsfs}%
\usepackage{xcolor}%
\usepackage{algorithm}%
\usepackage{algorithmicx}%
\usepackage{algpseudocode}%
\usepackage{listings}
\usepackage{tikz}
\usepackage{braket}
\usepackage{subcaption}
\begin{document}

\title{Toward Practical Quantum Machine Learning: A Novel Hybrid Quantum LSTM for Fraud Detection}

\author{\IEEEauthorblockN{Rushikesh Ubale}
\IEEEauthorblockA{\textit{Qkrishi Quantum}\\
India \\
rushikesh.ubale@qkrishi.com}
\and
\IEEEauthorblockN{Sujan K. K.}
\IEEEauthorblockA{\textit{Theoretical Sciences Unit}\\ \textit{Jawaharlal Nehru Centre For} \\
 \textit{Advanced Scientific Research}, India \\
sujanbharadwaj97@gmail.com}
\and
\IEEEauthorblockN{Sangram Deshpande}
\IEEEauthorblockA{\textit{Dept. of ECE} \\ \textit{NC State University}\\
Raleigh, NC, USA \\
ssdesh24@ncsu.edu}
\and
\IEEEauthorblockN{Gregory T. Byrd}
\IEEEauthorblockA{\textit{Dept. of ECE} \\ \textit{NC State University}\\
Raleigh, NC, USA \\
gbyrd@ncsu.edu}
}

\maketitle
\begin{abstract}
We present a novel hybrid quantum--classical neural network architecture for fraud detection that integrates a classical Long Short-Term Memory (LSTM) network with a variational quantum circuit. By leveraging quantum phenomena such as superposition and entanglement, our model enhances the feature representation of sequential transaction data, capturing complex non-linear patterns that are challenging for purely classical models. A comprehensive data preprocessing pipeline is employed to clean, encode, balance, and normalize a credit card fraud dataset, ensuring a fair comparison with baseline models. Notably, our hybrid approach achieves per-epoch training times in the range of 45--65 seconds, which is significantly faster than similar architectures reported in the literature, where training typically requires several minutes per epoch. Both classical and quantum gradients are jointly optimized via a unified backpropagation procedure employing the parameter-shift rule for the quantum parameters. Experimental evaluations demonstrate competitive improvements in accuracy, precision, recall, and F1 score relative to a conventional LSTM baseline. These results underscore the promise of hybrid quantum-classical techniques in advancing the efficiency and performance of fraud detection systems.
\end{abstract}


\section{Introduction}\label{sec1}

Access to the internet is growing globally, resulting in an increased number of online transactions. Along with the rise in total online transactions, online frauds are also increasing worldwide (provide supporting statistics). There are many categories of fraud, such as credit card fraud, online payment fraud, phishing fraud, and e-commerce fraud. Frauds can also be investment-related, healthcare-related, business-related, job-offer-related, and more. Machine Learning (ML) and Artificial Intelligence (AI) approaches have been employed extensively to tackle these issues. ML, in particular, has found significant applications in addressing fraud in sectors such as healthcare, e-commerce, insurance, credit card fraud, email-related fraud, and beyond. In this article, we focus on developing a model to address credit card fraud using Machine Learning techniques in conjunction with Quantum Computing.\\

Numerous models have been developed for credit card fraud detection (FD). Many of these are purely classical ML models, such as XGBoost, Random Forest (RF), Support Vector Machines (SVM), and deep learning models. Additionally, hybrid quantum-classical models, such as Quantum Support Vector Machines (QSVM), Quantum Convolutional Neural Networks (Quantum CNN), and Variational Quantum Classifiers (VQCs), have been explored. The integration of quantum principles into classical models offers unique advantages, such as leveraging entanglement, superposition, and access to complex spaces not achievable by classical models. Quantum models also demonstrate high expressivity, quantum parallelism, and the ability to represent data in higher-dimensional spaces via quantum feature maps, making them particularly effective. However, quantum computing faces challenges, including the interpretability of complex representations and the barren plateau problem. In this article, we leverage quantum principles in combination with classical LSTM models to address credit card fraud detection.\\

When addressing online fraudulent transactions, it is crucial to minimize false positives (FP), which lead to customer dissatisfaction, and false negatives (FN), which erode customers' trust in the banking system. Therefore, any ML model developed for this purpose must focus on reducing FP and FN (or related derived metrics). An effective model should identify important behavioral patterns and repeated sequences in the data to improve performance. Although there is no strict rule for feature selection, handling imbalanced datasets remains a significant challenge. The ratio of fraudulent to normal transactions is often very low, making it difficult to train accurate classification models. Techniques like SMOTE (Synthetic Minority Oversampling Technique) are commonly used to address this issue. Additionally, model robustness should be evaluated by testing on different datasets.

\section{Literature Review}

In recent years, fraud detection has attracted significant attention in both industry and academia due to its critical impact on financial systems. Researchers have explored a variety of approaches ranging from classical deep learning methods to emerging hybrid quantum-classical models. This section reviews the most relevant works, highlighting the motivation for our proposed model and emphasizing its architectural novelty.

\subsection{Classical Models}

Classical deep learning methods, particularly recurrent neural networks (RNNs) and their variants, have been widely applied to fraud detection. For example, Heryadi \textit{et al.} \cite{heryadi2017learning} compared Convolutional Neural Networks (CNNs), Stacked LSTMs, and CNN-LSTM hybrid models under varying normal-to-fraud ratios. Their study demonstrated that while Stacked LSTMs achieved higher training accuracy, CNNs provided better discrimination as measured by the area under the ROC curve (AUC), highlighting the trade-off between accuracy and discrimination capability.

Similarly, Guo \textit{et al.} \cite{guo2018learning} proposed the Historical Attention-based and Interactive LSTM (HAInt-LSTM), which introduces modifications to the LSTM forget gate to account for variable time intervals. By incorporating a self-historical attention mechanism, HAInt-LSTM better captures long-term behavioral patterns, thereby distinguishing normal from fraudulent activities more effectively than plain LSTM, GRU, phased-LSTM, or Update Gate RNN (UGRNN).

Other works, such as Cheng \textit{et al.} \cite{cheng2020spatio}, have integrated spatial and temporal features using models like the Spatial-Temporal Attention-based Neural Network (STAN). STAN combines 3D convolutions with spatio-temporal attention mechanisms to consider geographical and transactional timing aspects, outperforming classical models like Logistic Regression, Gradient Boosted Trees, and Deep Neural Networks.

Additionally, Alghofaili \textit{et al.} \cite{alghofaili2020financial} enhanced LSTM architectures by optimizing training time and accuracy. However, their reported dataset exhibited significant class imbalance, which limits the generalizability of their model's performance.

Recent enhancements have also involved integrating attention mechanisms with LSTMs. For instance, Benchaji \textit{et al.} \cite{benchaji2021enhanced} introduced an attention layer in conjunction with LSTM, leading to improvements in precision, recall, and overall accuracy compared to traditional models such as k-Nearest Neighbors (KNN), Artificial Neural Networks (ANN), and Support Vector Machines (SVM).

While these classical models have demonstrated success, they are inherently limited by the computational constraints of classical processing and their inability to efficiently handle high-dimensional feature spaces. This motivates the exploration of hybrid quantum-classical approaches to \textbf{overcome these limitations}.

\subsection{Hybrid Quantum-Classical Models}

Hybrid models that combine classical and quantum computing paradigms have recently emerged as promising alternatives, offering the potential to enhance the representational power of classical models through quantum encoding. Grossi \textit{et al.} \cite{grossi2022mixed} employed a hybrid quantum-classical model based on a Quantum Support Vector Machine (QSVM) and demonstrated improved performance by exploiting quantum-enhanced feature spaces. Their work illustrated that quantum kernels could provide superior classification accuracy compared to classical SVMs in high-dimensional spaces.

Similarly, Kyriienko \textit{et al.} \cite{kyriienko2022unsupervised} introduced an unsupervised quantum machine learning model that utilized quantum kernels to map classical features into an enlarged latent space, achieving higher expressivity compared to classical Radial Basis Function (RBF) kernels.

In the domain of fraud detection, Innan \textit{et al.} \cite{innan2024financial} applied Quantum Graph Neural Networks (Quantum GNN) to financial fraud detection, showing that quantum approaches can outperform classical counterparts in both accuracy and efficiency. Furthermore, Wang \textit{et al.} \cite{wang2022integrating} enhanced QSVMs with quantum annealing solvers, yielding improvements in both speed and predictive performance.

Notably, Innan \textit{et al.} \cite{innan2024qfnn} further pushed the boundaries by integrating federated learning with quantum neural networks, addressing data privacy concerns while achieving a remarkable accuracy of 0.95. This highlights the potential of hybrid models to balance performance with privacy preservation.

\subsection{Performance Comparison: Quantum vs Classical Models}

A particularly relevant study by Saad Zafar Khan \textit{et al.} \cite{khan2024quantum} compared classical and quantum LSTM models for solar power forecasting. Their findings indicated that the quantum model achieved better performance metrics. However, the authors reported an average training time per epoch of approximately 1 hour and 26 minutes. 

In our work, we achieve a significantly reduced epoch time of 45 to 65 seconds on a dataset with 10,000 data points while using PennyLane's \texttt{default.qubit} simulator, which runs on the CPU without GPU acceleration. This is noteworthy because the \texttt{default.qubit} simulator is a general-purpose state vector simulator, which is not specifically optimized for speed. In contrast, simulators such as \texttt{lightning.qubit} and \texttt{lightning.gpu} from PennyLane leverage advanced linear algebra libraries and GPU acceleration, enabling faster execution. 

Despite using the less optimized \texttt{default.qubit} simulator, our model achieves a considerably faster training time, indicating its efficiency in terms of computational overhead. This demonstrates the practical advantage of our hybrid quantum-classical architecture, as it offers both competitive accuracy and faster execution.

\subsection{Architectural Novelty and Design Choices}

\textbf{Our proposed model introduces several architectural innovations} that distinguish it from prior works, addressing their limitations in terms of expressivity and computational efficiency:

\begin{itemize}
    \item \textbf{Hybrid Architecture with Variational Quantum Circuits:} Unlike previous models that rely on either classical deep learning or shallow quantum models, we integrate a \textbf{variational quantum circuit (VQC)} with a classical LSTM. This hybrid design combines the temporal sequence modeling capabilities of LSTMs with the enhanced feature representation of quantum circuits, improving both accuracy and efficiency.

    \item \textbf{Efficient Quantum Encoding:} We use \textbf{AngleEmbedding} to efficiently map classical features into the quantum Hilbert space. This encoding technique allows the model to represent complex patterns using quantum rotations, increasing the \textbf{expressivity} of the model while maintaining low overhead.

    \item \textbf{Strongly Entangling Layers (SEL):} To enhance the feature representation power, we incorporate \textbf{Strongly Entangling Layers} in the VQC. These layers introduce controlled quantum entanglement between qubits, enabling the model to capture complex, non-linear dependencies in the data. Prior works have demonstrated that entanglement improves the expressive power of quantum models, making them more effective for high-dimensional problems \cite{Sim_2019}.

    \item \textbf{Efficient Training with CPU-Based Simulation:} Despite using the \texttt{default.qubit} simulator (CPU-based) instead of the more optimized \texttt{lightning.qubit} or \texttt{lightning.gpu} simulators, our model achieves \textbf{training times of 45–65 seconds per epoch} on 10,000 data points. This demonstrates that our architecture achieves high efficiency even without hardware acceleration, making it practical for real-world deployment.
\end{itemize}

\subsection{Motivation and Scope}

While prior studies have shown the effectiveness of hybrid quantum-classical models in improving accuracy, they often suffer from \textbf{computational inefficiencies} and \textbf{prolonged training times}. 
In contrast, our work addresses this limitation by demonstrating that a \textbf{carefully designed hybrid model} can achieve both \textbf{high accuracy} and \textbf{reduced training time}. 

Furthermore, our implementation uses the \texttt{default.qubit} simulator, which, despite being CPU-based, achieves efficient training times. This highlights the feasibility of deploying hybrid quantum-classical models on standard hardware without the immediate need for GPU acceleration, making our approach more accessible and practical for real-world applications. However, future studies could introduce GPU acceleration to further improve training speed and scale the model to larger datasets. Additionally, by training on a higher number of data samples, we can effectively mitigate overfitting, enhancing the model’s generalizability and reliability for real-world deployment.

\section{Data Preprocessing and Feature Engineering}

Our dataset consists of credit card transaction records that include various attributes such as transaction timestamps, geospatial coordinates, and customer demographics. To prepare the data for robust fraud detection, we performed extensive preprocessing and feature engineering. These steps transform raw data into informative features while addressing issues such as class imbalance.

\subsection{Temporal and Geospatial Feature Extraction}

First, the raw transaction timestamp (\texttt{trans\_date\_trans\_time}) was converted into a standardized \texttt{datetime} format. From this, we extracted several time-based features (hour, day, weekday, month, and year) to capture temporal patterns that may be indicative of fraudulent behavior. In addition, we derived a key geospatial feature: the distance between customer and merchant locations. To calculate this, we used the well-known haversine formula, which computes the great-circle distance between two points on a sphere. This is especially useful for detecting anomalies when transactions occur over unexpectedly large distances. The formula is given by:
\[
d = 2R \cdot \arctan2\!\left(\sqrt{a}, \sqrt{1-a}\right),
\]
with
\[
a = \sin^2\!\left(\frac{\Delta \phi}{2}\right) + \cos(\phi_1) \cos(\phi_2) \sin^2\!\left(\frac{\Delta \lambda}{2}\right).
\]
Here, \(R\) is the Earth's radius (6371 km), \(\phi_1\) and \(\phi_2\) are the latitudes (in radians) of the customer and merchant respectively, and \(\Delta \lambda\) is the difference in their longitudes (in radians). Including this feature is necessary to quantify the spatial relationship between transaction parties, which can be critical for detecting geographically inconsistent or suspicious transactions.

\subsection{Categorical Encoding and Demographic Feature Engineering}

Categorical variables such as \texttt{category}, \texttt{gender}, \texttt{state}, and \texttt{job} were transformed into numerical values using label encoding. This conversion is essential for incorporating non-numeric information into machine learning algorithms. Additionally, we computed the customer age from the date of birth, which serves as a valuable demographic indicator for assessing fraud risk.

\subsection{Balancing the Dataset}

Fraud detection datasets often suffer from severe class imbalance, where non-fraudulent transactions greatly outnumber fraudulent ones. To mitigate this issue, we applied resampling techniques: oversampling the minority (fraud) class and downsampling the majority (non-fraud) class to achieve a balanced dataset of 50,000 samples per class. This balance is crucial for preventing the predictive model from being biased toward the majority class and for ensuring that both classes are adequately represented during model training.

Overall, these preprocessing and feature engineering steps were necessary to convert heterogeneous raw data into a structured, refined dataset. By capturing temporal, spatial, and categorical nuances, we enhance the interpretability and performance of our fraud detection models.

\section{Methodology}

In this work, we propose a hybrid quantum-classical neural network architecture for fraud detection, integrating a Long Short-Term Memory (LSTM) network for temporal feature extraction with a quantum layer for enhanced representation learning. The methodology consists of several stages: data preprocessing, classical feature extraction, quantum embedding, and final classification. First, raw transaction data undergoes preprocessing, including normalization and encoding, to ensure compatibility with the neural network input. The LSTM processes sequential features, capturing temporal dependencies, and its output is mapped to a lower-dimensional space using a fully connected (FC) layer. This reduced feature set is then passed to a variational quantum circuit (VQC), where quantum embeddings and entangling layers transform the data. The quantum circuit’s measurement outcomes are subsequently processed by an additional FC layer, producing the final classification output. The entire pipeline is trained using backpropagation, with classical gradients propagating through the quantum circuit using the parameter-shift rule. The following sections detail each stage, including data preparation, model architecture, training setup, and evaluation metrics.

\subsection{Data Acquisition and Preprocessing}

The dataset used in this study is a synthetically generated credit card fraud dataset designed to mimic real-world transaction data. It contains 27 features, categorized into different types, as shown in Table~\ref{tab:data_features}. These features include card details, merchant information, transaction amount, personal identifiers, geographical data, transaction metadata, and additional encoded categorical variables. 

As described in Section 2, the raw data was initially loaded, and basic cleaning was performed. In this section, we detail the subsequent preprocessing steps applied prior to model training.

\begin{table*}[h]
    \centering
    \caption{Feature Description of the Credit Card Fraud Dataset}
    \label{tab:data_features}
    \begin{tabular}{|l|l|p{8cm}|}
        \hline
        \textbf{Feature Category} & \textbf{Feature Name} & \textbf{Description} \\ \hline
        \multirow{2}{*}{Card Details} & \texttt{cc\_num} & Credit card number (hashed) \\ 
        & \texttt{merchant} & Merchant information for transaction \\ \hline
        Transaction Amount & \texttt{amt} & Transaction amount in USD \\ \hline
        \multirow{4}{*}{Personal Identifiers} & \texttt{first}, \texttt{last} & First and last name of cardholder \\ 
        & \texttt{street} & Street address of cardholder \\ 
        & \texttt{gender\_encoded} & Encoded gender information \\ 
        & \texttt{customer\_age} & Estimated age of the customer \\ \hline
        \multirow{5}{*}{Geographical Data} & \texttt{city}, \texttt{zip} & City and ZIP code of the transaction \\ 
        & \texttt{lat}, \texttt{long} & Latitude and longitude coordinates \\ 
        & \texttt{city\_pop} & Population of the city \\ 
        & \texttt{state\_encoded} & Encoded state information \\ \hline
        \multirow{5}{*}{Transaction Metadata} & \texttt{unix\_time} & Timestamp of the transaction \\ 
        & \texttt{transaction\_hour} & Hour of the transaction \\ 
        & \texttt{transaction\_day} & Day of the transaction \\ 
        & \texttt{transaction\_weekday} & Day of the week \\ 
        & \texttt{transaction\_month}, \texttt{transaction\_year} & Month and year of transaction \\ \hline
        \multirow{3}{*}{Additional Features} & \texttt{category\_encoded} & Encoded transaction category \\ 
        & \texttt{job\_encoded} & Encoded job category of the cardholder \\ 
        & \texttt{customer-merchant distance} & Calculated distance between customer and merchant \\ \hline
    \end{tabular}
\end{table*}

\subsubsection{Data Acquisition and Preprocessing Pipeline}

The data acquisition and preprocessing pipeline consists of the following steps to prepare the dataset for model training and evaluation:

\begin{enumerate}
    \item \textbf{Data Cleaning:}  
    To reduce irrelevant information, personally identifiable columns such as \texttt{first}, \texttt{last}, \texttt{street}, and \texttt{trans\_num} are removed. This step eliminates unnecessary attributes that do not contribute to fraud detection while complying with data protection regulations.
    
    \item \textbf{Categorical Encoding:}  
    Since machine learning models require numerical inputs, categorical features such as \texttt{merchant} and \texttt{city} are transformed using \textit{label encoding}. Each unique category is assigned a numerical value, allowing models to process these features effectively.
    
    \item \textbf{Dataset Subsetting:}  
    Due to computational constraints, we select a balanced subset of 10,000 transactions, consisting of 5,000 fraudulent and 5,000 non-fraudulent transactions. This ensures that our dataset remains manageable while maintaining class balance, preventing model bias.

    \item \textbf{Feature Normalization:}  
    To standardize continuous numerical features, we apply \textit{z-score normalization}. Each feature $x$ is normalized as follows:
    \[
    x_{\text{norm}} = \frac{x - \mu}{\sigma},
    \]
    where $\mu$ and $\sigma$ represent the mean and standard deviation of the feature, respectively. This transformation ensures that all numerical attributes are on a comparable scale, improving model convergence and performance.

    \item \textbf{Data Partitioning:}  
    The preprocessed dataset is split into three subsets: training (70\%), validation (15\%), and test (15\%). \textit{Stratified sampling} is used to maintain an equal proportion of fraudulent and non-fraudulent transactions in each subset, ensuring that the model learns effectively from both classes.

    \item \textbf{Tensor Conversion:}  
    To enable efficient computation in deep learning models, the processed data is converted from NumPy arrays to PyTorch tensors. This step facilitates seamless integration with PyTorch-based architectures, optimizing performance during training and inference.
\end{enumerate}

\subsection{Model Architecture}

Our hybrid quantum--classical neural network is composed of three major components: 
(1) a classical Long Short-Term Memory (LSTM) module for extracting temporal features, 
(2) a dimensionality-reduction layer mapping the LSTM state into a dimension suitable for quantum processing, 
and (3) a variational quantum circuit (VQC) followed by a final fully connected (FC) layer for classification. 
Figure~\ref{fig:quantum_circuit} illustrates the quantum portion of the architecture, highlighting how the 
classical data is embedded, transformed through strongly entangling layers, and measured.

\subsubsection{Classical LSTM Component}
The LSTM \cite{Hochreiter1997} processes sequential transaction data to capture long-term dependencies, which 
is crucial for fraud detection. Let $x_t$ be the input at time $t$, and let $h_t$ and $C_t$ denote the hidden 
and cell states, respectively. The LSTM computes:
\begin{align}
    i_t &= \sigma(W_i x_t + U_i h_{t-1} + b_i), \label{eq:input_gate}\\
    f_t &= \sigma(W_f x_t + U_f h_{t-1} + b_f), \label{eq:forget_gate}\\
    \tilde{C}_t &= \tanh(W_C x_t + U_C h_{t-1} + b_C), \label{eq:cell_candidate}\\
    C_t &= f_t \odot C_{t-1} + i_t \odot \tilde{C}_t, \label{eq:cell_state}\\
    o_t &= \sigma(W_o x_t + U_o h_{t-1} + b_o), \label{eq:output_gate}\\
    h_t &= o_t \odot \tanh(C_t), \label{eq:hidden_state}
\end{align}
where $\sigma(\cdot)$ is the sigmoid activation function, $\tanh(\cdot)$ is the hyperbolic tangent, $W$ and $U$ 
are weight matrices, $b$ is the bias term, and $\odot$ denotes element-wise multiplication. 
Equations~\eqref{eq:input_gate}--\eqref{eq:hidden_state} describe how the LSTM selectively remembers or forgets 
information at each time step \cite{Goodfellow2016, Chen2022LSTM}.
\subsubsection{Handling of Time and Batch Dimensions}
In our implementation, the LSTM is configured with \texttt{batch\_first=True} and expects input tensors of shape 
\[
(\text{batch\_size}, \text{sequence\_length}, \text{input\_size}).
\]
Since each transaction is represented by a static feature vector (i.e., with shape 
\((\text{batch\_size}, \text{input\_size})\)), we insert a dummy time dimension using \texttt{unsqueeze(1)}. 
This reshapes the input to 
\[
(\text{batch\_size}, 1, \text{input\_size}),
\]
effectively treating each sample as a sequence of length 1. The LSTM processes this input and outputs the 
hidden state corresponding to this single time step, which is then used for subsequent dimensionality reduction 
and quantum embedding.

\begin{figure}[ht]
    \centering
    \includegraphics[width=0.6\linewidth]{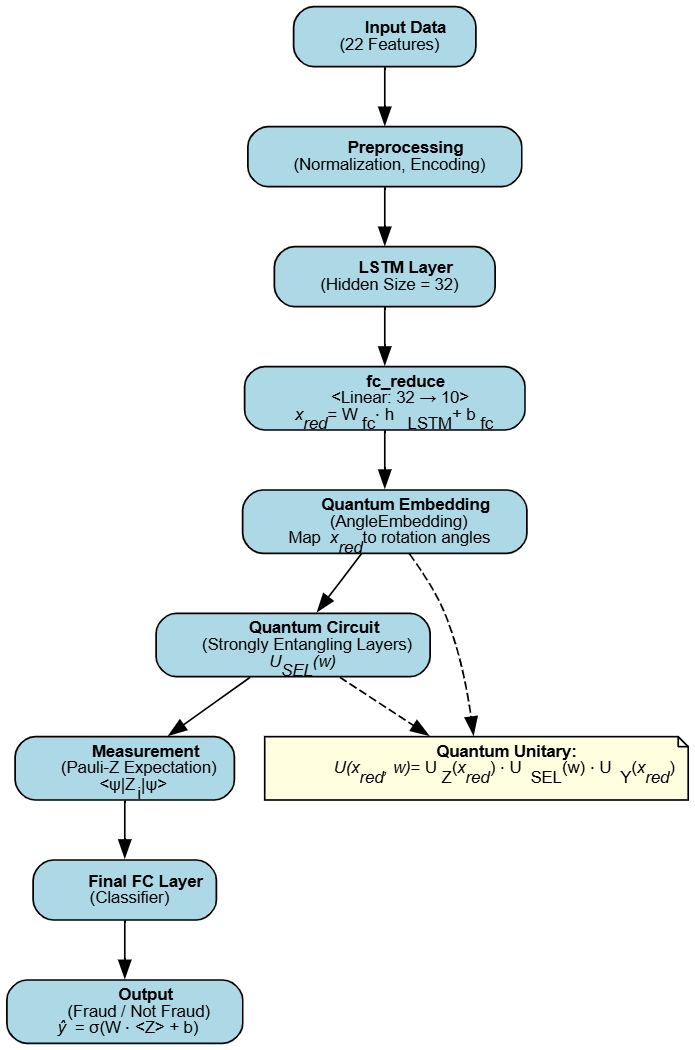}
    \caption{Model Architecture}
    \label{fig:enter-label}
\end{figure}
\subsubsection{Dimensionality Reduction via FC Layer}
After processing all time steps, the final hidden state $h_T$ is extracted. Often, $h_T$ has a higher dimensionality 
than the number of qubits in our quantum circuit. We therefore apply a fully connected (FC) layer that projects 
$h_T$ onto $\mathbb{R}^{n_{\text{qubits}}}$. This ensures that the classical feature vector is appropriately 
sized for the subsequent quantum embedding stage.

\subsubsection{Quantum Layer}
The primary innovation lies in the quantum layer, implemented as a variational quantum circuit (VQC). 
This layer exploits quantum effects---such as superposition and entanglement---to enhance feature representation 
\cite{Biamonte2017, Schuld2019}.

\paragraph{Angle Embedding and Strongly Entangling Layers.}
Figure~\ref{fig:quantum_circuit} depicts 10 qubit wires labeled from 0 to 9. Initially, the classical feature 
vector $\vec{x} = [x_1, x_2, \dots, x_{n_{\text{qubits}}}]$ is encoded onto these qubits via \emph{AngleEmbedding}:
\begin{equation}
|\psi(\vec{x})\rangle \;=\; \bigotimes_{i=1}^{n_{\text{qubits}}} R_Y(x_i)\,\ket{0},
\end{equation}
where $R_Y(x_i)$ is a rotation about the $Y$ axis by angle $x_i$, mapping the classical data into the quantum 
Hilbert space \cite{Havlicek2019, Schuld2019}. Next, \emph{StronglyEntanglingLayers} apply repeated blocks of 
entangling gates (e.g., CNOT) interspersed with single-qubit parameterized rotations:
\begin{equation}
U(\theta) \;=\; \prod_{\ell=1}^{L} U_{\text{entangle}}^{(\ell)}(\theta^{(\ell)}),
\end{equation}
where $\theta = \{\theta^{(1)}, \theta^{(2)}, \dots, \theta^{(L)}\}$ are trainable parameters. These layers 
introduce correlations among qubits, enabling the circuit to learn highly non-linear mappings of the input 
features \cite{Mitarai2018, Bergholm2018}.

\paragraph{Measurement.}
After the final entangling layer, the circuit measures each qubit in the Pauli-$Z$ basis:
\begin{equation}
q_i \;=\; \langle \psi(\vec{x}, \theta)|\,Z_i\,|\psi(\vec{x}, \theta)\rangle,\quad i=1,\dots,n_{\text{qubits}}.
\end{equation}
This yields a real-valued vector $\vec{q} = [q_1, q_2, \dots, q_{n_{\text{qubits}}}]$ that captures quantum-enhanced 
features of the data. By leveraging entanglement and superposition, the circuit can represent transformations 
that may be difficult for purely classical architectures \cite{Chen2022Quantum}.

\begin{figure}
    \centering
    \includegraphics[width=0.99\linewidth]{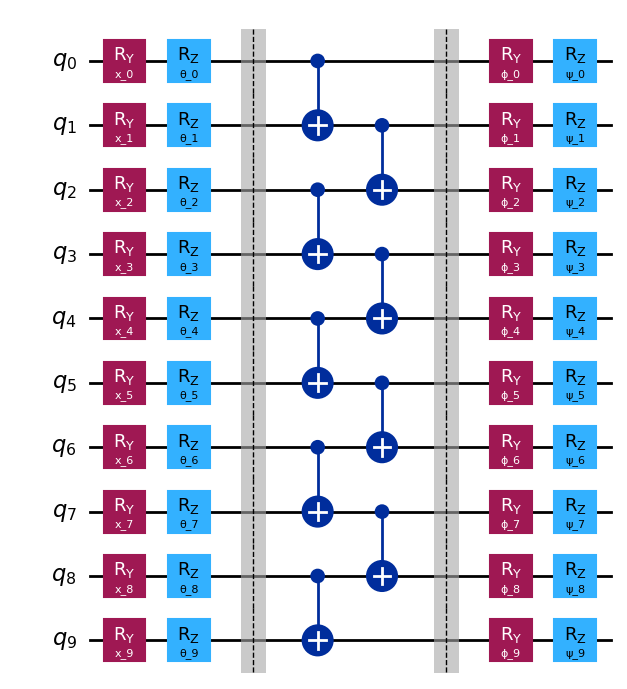}
    \caption{Decomposed quantum circuit used in our QLSTM-based fraud detection model. Each qubit is initialized in the state $\ket{0}$ and undergoes a data-encoding $R_y$ rotation with parameter $x_i$, followed by a trainable $R_z(\theta_i)$. The circuit includes barriers to demarcate the encoding, entanglement, and final transformation stages for clarity. The entanglement layer is implemented using staggered CNOT gates between adjacent qubits to ensure local entanglement. This is followed by a second set of trainable rotations $R_y(\phi_i)$ and $R_z(\psi_i)$. All qubits are measured in the computational (Z) basis. This diagram is a simplified decomposition of the PennyLane \texttt{StronglyEntanglingLayers} template used in the model.}
    \label{fig:enter-label}
\end{figure}
\footnotetext{This diagram is a simplified decomposition of the Pennylane \texttt{StronglyEntanglingLayers} template used in the model. The actual implementation may include additional gate-level optimizations not shown here.}
\subsubsection{Final Classification Layer}
The quantum output $\vec{q}$ is passed into a final fully connected layer:
\begin{equation}
\hat{y} \;=\; \sigma\Bigl(W_{\text{out}} \,\vec{q} + b_{\text{out}}\Bigr),
\end{equation}
where $\sigma(\cdot)$ is the sigmoid function, $W_{\text{out}}$ is the weight matrix, and $b_{\text{out}}$ 
is the bias term. The resulting scalar $\hat{y}$ represents the predicted probability of fraud, with values 
closer to 1 indicating a higher likelihood of fraudulent activity.

\subsubsection{Note on Parameter Optimization}

\paragraph{Integration of Classical and Quantum Gradients.}
Both the classical (LSTM and FC layers) and quantum circuit parameters are trained end-to-end in a unified 
backpropagation loop \cite{Goodfellow2016, Bergholm2018}. Concretely, the partial derivatives of the loss 
function with respect to the classical parameters (e.g., $W_i, U_i, b_i$ in the LSTM) are computed via standard 
automatic differentiation. However, gradients for the quantum parameters $\theta$ require the \textit{parameter-shift rule}:
\begin{enumerate}
    \item \textbf{Forward Pass:} The model processes the input sequence through the LSTM and FC layer, 
    then encodes the data into the quantum circuit. After quantum operations, a measurement vector 
    $\vec{q}$ is produced, which is fed into the final FC layer to obtain the loss $\mathcal{L}$.
    
    \item \textbf{Quantum Parameter-Shift:} For each parameter $\theta_j$, the circuit is evaluated 
    twice: once at $\theta_j + \Delta$ and once at $\theta_j - \Delta$, where $\Delta$ is a small shift 
    (often $\Delta = \pi/2$). The partial derivative is approximated by:
    \[
    \frac{\partial \mathcal{L}}{\partial \theta_j} \;\approx\; \frac{\mathcal{L}(\theta_j + \Delta) \;-\; \mathcal{L}(\theta_j - \Delta)}{2\,\sin(\Delta)}.
    \]
    
    \item \textbf{Combine Gradients:} The gradient contributions from the quantum circuit parameters 
    are then merged with those from the classical LSTM and FC layers. Because all components are part 
    of the same computational graph, the chain rule naturally propagates errors back through both 
    classical and quantum modules.
    
    \item \textbf{Optimizer Step:} A single optimizer (e.g., Adam) applies the combined gradient updates 
    to all parameters in one pass, ensuring consistent learning dynamics across the entire model.
\end{enumerate}
This unified approach allows classical layers to learn temporal features while the quantum circuit simultaneously 
learns an expressive mapping in Hilbert space, resulting in a more powerful end-to-end system \cite{Mitarai2018, 
Chen2022Quantum}.

\paragraph{Practical Implementation.}
We implement this architecture using the PennyLane library \cite{Bergholm2018} for quantum operations and 
PyTorch for classical layers. PennyLane seamlessly integrates with PyTorch, allowing us to compute classical 
gradients using automatic differentiation and quantum gradients via the parameter-shift rule in a single 
backpropagation loop.

\noindent
\textbf{Overall Pipeline:}  
\begin{enumerate}
    \item \textbf{Sequential Modeling:} The LSTM extracts temporal features from transaction sequences.
    \item \textbf{Dimension Matching:} An FC layer reduces $h_T$ to size $n_{\text{qubits}}$.
    \item \textbf{Quantum Circuit:} The quantum layer embeds and entangles these features, outputting a 
    measurement vector $\vec{q}$.
    \item \textbf{Classification:} The final FC layer outputs a probability of fraud, with all parameters 
    updated end-to-end in one training loop.
\end{enumerate}
This design exploits both classical time-series modeling and quantum feature representation to enhance fraud 
detection performance \cite{Biamonte2017, Havlicek2019, Schuld2019, Chen2022Quantum}.

\section{Training and Evaluation}

After defining our hybrid quantum-classical model architecture, we proceed with the training procedure, hyperparameter configuration, and evaluation protocol. The following subsections describe how the model is optimized, validated, and tested, as well as the metrics used to assess performance.
\subsubsection*{Flow Diagram of the Training Process}
\begin{figure}[ht]
    \centering
    \begin{tikzpicture}[node distance=1.8cm, auto, >=stealth, thick]
        \tikzstyle{startstop} = [rectangle, rounded corners, draw=black, fill=red!30, text width=7em, text centered, minimum height=2em]
        \tikzstyle{process} = [rectangle, draw=black, fill=blue!20, text width=10em, text centered, minimum height=2em]
        \tikzstyle{decision} = [diamond, draw=black, fill=green!30, text width=7em, text centered, aspect=2]
        \tikzstyle{arrow} = [thick,->,>=stealth]
        
        \node (data) [startstop] {Data Loading};
        \node (batch) [process, below of=data] {Batch Processing (DataLoader)};
        \node (forward) [process, below of=batch] {Forward Pass \\ (LSTM $\rightarrow$ FC $\rightarrow$ Quantum Layer)};
        \node (grad) [process, below of=forward] {Gradient Computation \\ (Classical + Quantum)};
        \node (update) [process, below of=grad] {Optimizer Update (Adam)};
        \node (log) [process, below of=update] {Logging Metrics};
        
        \draw [arrow] (data) -- (batch);
        \draw [arrow] (batch) -- (forward);
        \draw [arrow] (forward) -- (grad);
        \draw [arrow] (grad) -- (update);
        \draw [arrow] (update) -- (log);
    \end{tikzpicture}
    \caption{Flow diagram summarizing the training process: data loading, forward pass through the hybrid model, gradient evaluation (integrating classical gradients and quantum gradients via the parameter-shift rule as detailed in Methodology Section), and optimizer update followed by metric logging.}
    \label{fig:training_flow}
\end{figure}
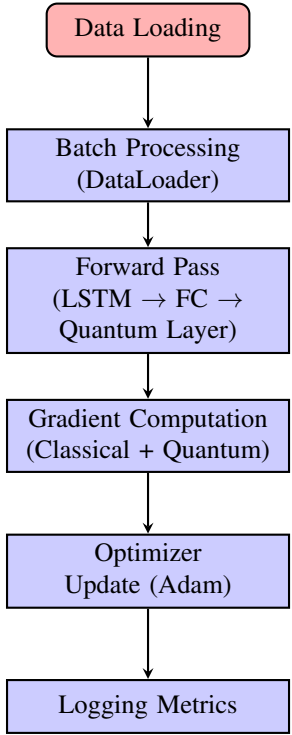

Figure~\ref{fig:training_flow} summarizes the training pipeline, outlining the sequence of steps from data loading to optimizer updates.

\subsection{Training Procedure}

In this section, we describe the complete training process for our hybrid quantum--classical model. We detail the configuration of data loaders and batch processing, the choice of loss function and optimizer, the integrated gradient computation for both classical and quantum parameters, and our logging strategy over the training epochs.

\subsubsection{ Data Loaders and Batch Processing}
To ensure efficient and stable training, we employ PyTorch's \texttt{DataLoader} for batch processing of the dataset. The data is split into training, validation, and test sets with stratified sampling to maintain class balance. For training, we set the batch size to 32 and enable shuffling:
\begin{itemize}
    \item \textbf{Batch Size:} A batch size of 32 is chosen to balance between memory efficiency and convergence speed. Smaller batches can introduce noisy gradient estimates, while larger batches may require more memory and lead to slower updates.
    \item \textbf{Shuffling:} Shuffling the data at each epoch ensures that the model does not learn spurious order-related patterns and improves the stability of the training process.
\end{itemize}
These steps help mitigate overfitting and improve the overall convergence of the model.

\subsubsection{ Loss Function and Optimizer}
For binary classification tasks such as fraud detection, we use the \texttt{BCEWithLogitsLoss} function provided by PyTorch. This loss function combines the sigmoid activation and the binary cross entropy loss into one stable computation:
\[
\mathcal{L}(y, \hat{y}) = -\frac{1}{N}\sum_{i=1}^{N}\Bigl[y_i \log\bigl(\sigma(\hat{y}_i)\bigr) + (1-y_i)\log\bigl(1-\sigma(\hat{y}_i)\bigr)\Bigr],
\]
where \(\sigma(\cdot)\) denotes the sigmoid function and \(N\) is the number of samples in the batch \cite{Goodfellow2016}. 

We use the Adam optimizer for its adaptive learning rate properties, which is especially useful when training deep networks with both classical and quantum components. The hyperparameters are set as follows:
\begin{itemize}
    \item \textbf{Learning Rate:} \( \eta = 0.001 \). This rate offers a good balance between convergence speed and the risk of overshooting minima.
    \item \textbf{Weight Decay:} \(1 \times 10^{-4}\) to regularize the model and prevent overfitting.
\end{itemize}
The hyperparameters for our model were chosen through a combination of insights from prior literature and empirical tuning on a validation set. A batch size of 32 was selected as it provided a good trade-off between gradient noise reduction and computational efficiency while fitting within available hardware memory. Similarly, the learning rate of 0.001 was chosen based on recommendations from previous studies \cite{Kingma2014} and further refined via grid search to ensure stable convergence. A weight decay of $1\times10^{-4}$ was incorporated to mitigate overfitting, in line with standard regularization practices in deep learning \cite{Goodfellow2016}. These choices have been validated across multiple runs and are reported here to facilitate reproducibility of our approach.

\subsubsection{Gradient Computation (Classical + Quantum)}
Our model integrates both classical layers (LSTM, FC layers) and a quantum layer. For classical layers, gradients are computed automatically via PyTorch's autograd engine. However, gradients for the quantum circuit parameters require a different approach known as the \textit{parameter-shift rule} \cite{Mitarai2018, Bergholm2018}:
\begin{enumerate}
    \item \textbf{Forward Pass:} The model processes the input data through the LSTM and FC layers, encodes it into the quantum circuit, and obtains the measurement vector \(\vec{q}\). This is then passed to the final FC layer to compute the loss \(\mathcal{L}\).
    \item \textbf{Quantum Gradient Evaluation:} 
Our training process computes gradients for both the classical layers (LSTM and FC layers) and the quantum circuit in a unified backpropagation loop. For the classical components, PyTorch's automatic differentiation efficiently computes the gradients. In contrast, the quantum circuit parameters are updated using the parameter-shift rule, as described in Methodology Section. Briefly, this rule estimates the gradient for each quantum parameter by perturbing it by a small shift (typically \(\Delta = \pi/2\)) and measuring the resulting change in the loss. By integrating these quantum gradients with the classical ones via the chain rule, we obtain a complete gradient vector for all parameters. Additionally, we employ standard techniques such as gradient clipping (to prevent exploding gradients in the LSTM) and learning rate scheduling to enhance training stability.

    \item \textbf{Gradient Integration:} The gradients from the quantum circuit are then merged with those from the classical layers using the chain rule. This unified gradient is used to update all parameters simultaneously in a single optimizer step.
\end{enumerate}
Additionally, we implement gradient clipping for the LSTM layers to avoid exploding gradients, and we may use learning rate scheduling to adjust the learning rate dynamically during training.

\subsubsection{Epochs and Logging}
We train our model for a total of 80 epochs. At each epoch, we track the following metrics:
\begin{itemize}
    \item \textbf{Training Loss:} Computed over all batches in the training set.
    \item \textbf{Validation Loss and Accuracy:} Evaluated on the validation set at the end of each epoch to monitor overfitting and convergence.
    \item \textbf{Epoch Time:} The duration of each epoch is recorded to assess training efficiency.
\end{itemize}
We log these metrics and print summary statistics after each epoch to monitor training progress. Early-stopping criteria are not enforced in the present experiments, but checkpoint saving is implemented to allow recovery of the best-performing model based on validation loss. This logging strategy ensures that we can analyze the training dynamics and adjust hyperparameters if needed.

\section{Benchmarking Models}

To assess the effectiveness of the quantum-enhanced model, we benchmark its performance against a well-established classical deep learning model: the \textbf{Classical LSTM}. This ensures that improvements (if any) are attributable to quantum advantages rather than architectural disparities.

\subsection{Classical LSTM Model}

Long Short-Term Memory (LSTM) networks \cite{Hochreiter1997} are a type of recurrent neural network (RNN) designed to capture long-range dependencies in sequential data. The classical LSTM model in this benchmark closely mirrors the quantum model’s data processing pipeline, ensuring a fair comparison.

\subsubsection{Data Preprocessing and Setup}

The dataset used for training both the Classical LSTM and Quantum LSTM models originates from a balanced credit card fraud detection dataset. The preprocessing steps are identical in both cases:
\begin{itemize}
    \item Encoding categorical variables using \texttt{LabelEncoder}.
    \item Standardizing numerical features with \texttt{StandardScaler}.
    \item Stratified train-validation-test split to maintain class balance.
\end{itemize}

The processed data is then wrapped into PyTorch \texttt{Dataset} and \texttt{DataLoader} objects for efficient mini-batch training.

\subsubsection{Classical LSTM Architecture}

The classical LSTM model follows a two-layer LSTM configuration with a dropout layer for regularization. The architecture consists of:
\begin{itemize}
    \item An \textbf{LSTM} module with \texttt{hidden\_size} of 32 and 2 layers.
    \item A fully connected (\texttt{fc}) layer mapping the final LSTM output to a single scalar prediction.
    \item A dropout rate of 0.3 to prevent overfitting.
\end{itemize}

The model is trained using the \textbf{Binary Cross-Entropy Loss with Logits} (\texttt{BCEWithLogitsLoss}) since fraud detection is a binary classification problem. The optimizer chosen is \textbf{Adam} with a learning rate of 0.001 and weight decay of $10^{-4}$.

\subsubsection{Training and Evaluation}

The training follows an iterative mini-batch gradient descent approach:
\begin{itemize}
    \item Each batch is forward-passed through the LSTM, and loss is computed.
    \item Gradients are backpropagated, and weights are updated via Adam optimization.
    \item Training is conducted for \textbf{80 epochs}.
\end{itemize}

Validation is performed at each epoch to monitor generalization. The final test evaluation computes:
\begin{itemize}
    \item \textbf{Accuracy}
    \item \textbf{Precision, Recall, and F1-score}
    \item \textbf{Confusion Matrix}
    \item \textbf{Inference Time}
\end{itemize}

\subsection{Comparison with Quantum LSTM}

Both the Classical and Quantum LSTM models share the following key characteristics:
\begin{itemize}
    \item \textbf{Data Processing:} Identical feature engineering and normalization steps.
    \item \textbf{Batch-based Training:} Both models utilize PyTorch's DataLoader for efficient gradient computation.
    \item \textbf{LSTM Core:} The Classical LSTM uses standard recurrent gates, whereas the Quantum LSTM replaces some transformations with variational quantum circuits.
    \item \textbf{Loss Function and Optimization:} Both use \texttt{BCEWithLogitsLoss} and Adam optimizer.
\end{itemize}

However, the primary distinction lies in the representation of sequential dependencies:
\begin{itemize}
    \item \textbf{Classical LSTM:} Uses standard matrix multiplications and nonlinear activations.
    \item \textbf{Quantum LSTM:} Incorporates quantum-parameterized transformations to enhance expressivity.
\end{itemize}

This direct comparison ensures that any observed improvements in the Quantum LSTM are due to quantum advantages rather than differences in data handling, model structure, or training strategy.

\section{Results}
\label{sec:results}

In this section, we present the experimental outcomes of our hybrid quantum-classical model for fraud detection. We begin by analyzing the training and validation dynamics, then evaluate the final performance on the test set. We also discuss the computational efficiency of our approach by examining the per-epoch training time.
\subsection{Quantum Hybrid Model Results}
\subsubsection{Training and Validation}

\subsubsection{Loss and Accuracy Curves}
\begin{figure}[ht]
    \centering
    \begin{subfigure}[b]{0.48\linewidth}
        \centering
        \includegraphics[width=\linewidth]{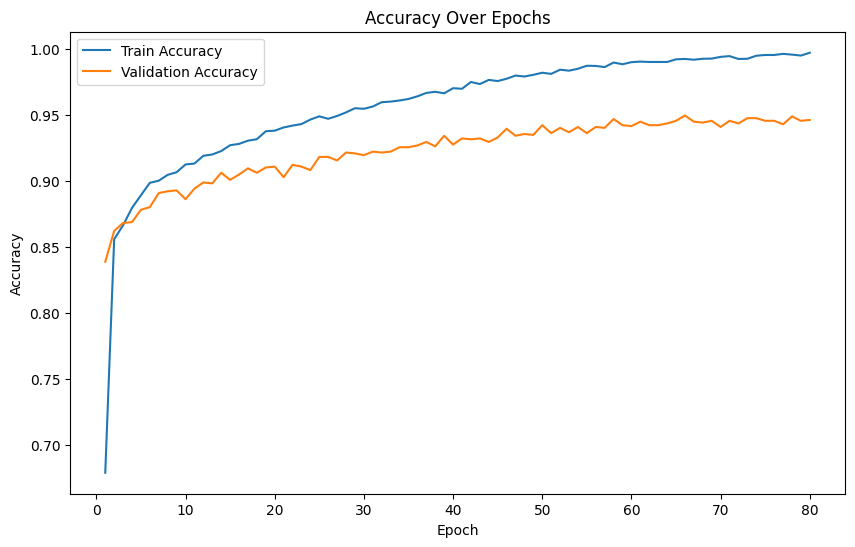} 
        \caption{Training and Validation Accuracy}
        \label{fig:accuracy_sub}
    \end{subfigure}
    \hfill
    \begin{subfigure}[b]{0.48\linewidth}
        \centering
        \includegraphics[width=\linewidth]{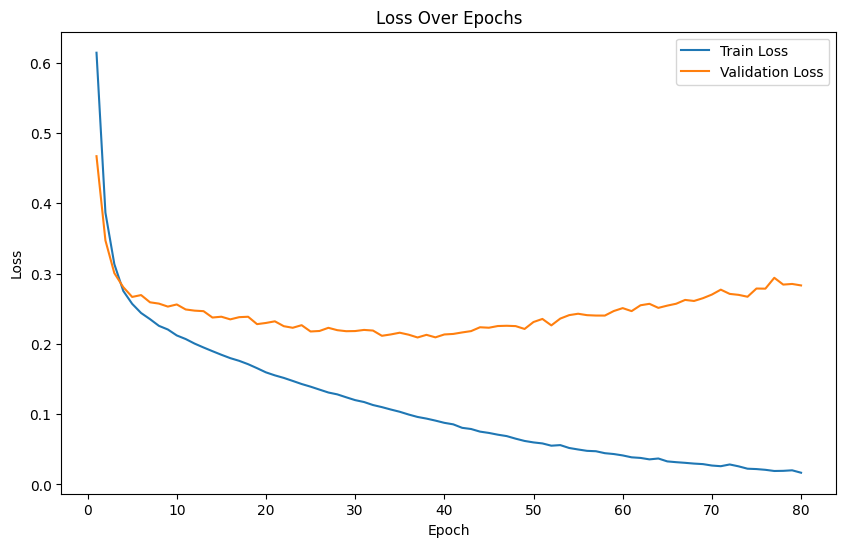} 
        \caption{Training and Validation Loss}
        \label{fig:loss_sub}
    \end{subfigure}
    \caption{Combined training performance over 80 epochs. The left subfigure shows accuracy, while the right subfigure shows loss.}
    \label{fig:combined_training}
\end{figure}

Figure~\ref{fig:accuracy_sub} illustrates the \textit{training and validation accuracy} throughout the same training period. The training accuracy rises quickly from about 70\% to over 95\%, eventually approaching 99\% by epoch 80. The validation accuracy follows a similar pattern, reaching around 95\%. This convergence trend indicates that the model successfully captures the temporal and feature-level correlations in the fraud dataset.
Figure~\ref{fig:loss_sub} shows the \textit{training and validation loss} over 80 epochs. Initially, the training loss starts around 0.6, dropping significantly during the first 10 epochs and then steadily declining until the end of training. The validation loss also decreases, indicating that the model is learning effectively without severe overfitting. 

\subsubsection{Time Per Epoch}

\begin{figure}[ht]
    \centering
    \includegraphics[width=0.5\linewidth]{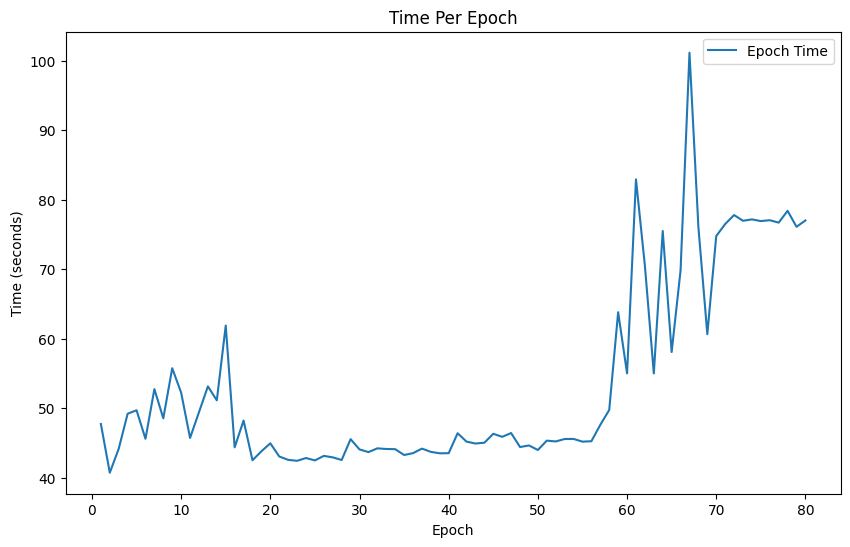} 
    \caption{Time Per Epoch for Training. Most epochs require 45--65 seconds, showcasing the model's relatively fast convergence despite incorporating quantum layers.}
    \label{fig:time_curve}
\end{figure}

Beyond predictive performance, computational efficiency is a key factor in hybrid quantum models. Figure~\ref{fig:time_curve} plots the time (in seconds) taken per epoch. On average, our model requires 45--65 seconds per epoch, with occasional spikes. Compared to other hybrid quantum LSTM approaches reported in the literature (often requiring several minutes or even hours per epoch), our model demonstrates a notable improvement in efficiency. This speedup stems from an optimized quantum circuit design and efficient integration of quantum and classical gradients.

\subsubsection{Test Evaluation and Analysis}

After completing the 80-epoch training phase, we evaluated the model on a held-out test set. The final performance metrics are as follows:
\begin{itemize}
    \item \textbf{Accuracy:} 95.33\%
    \item \textbf{Precision:} 94.16\%
    \item \textbf{Recall:} 96.67\%
    \item \textbf{F1 Score:} 95.39\%
    \item \textbf{Inference Time on Test Set:} 6.32 seconds
\end{itemize}
These metrics demonstrate that the hybrid quantum--classical approach excels in classifying both fraudulent and non-fraudulent transactions, achieving a balance between high recall (critical for catching fraud) and strong precision (to minimize false alarms).

\begin{figure}[ht]
    \centering
    \includegraphics[width=0.6\linewidth]{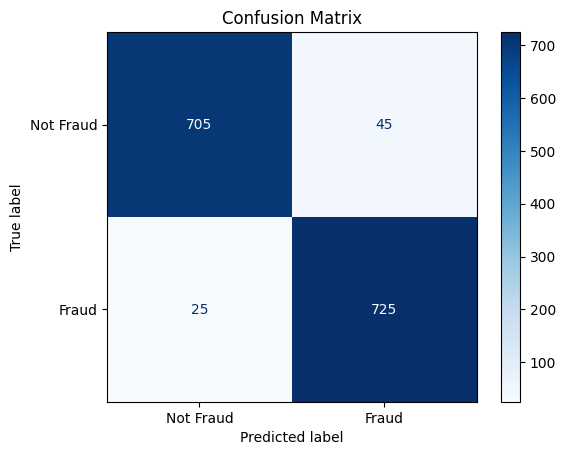} 
    \caption{Confusion Matrix on the Test Set. The model correctly identifies the majority of both fraudulent and non-fraudulent transactions.}
    \label{fig:conf_matrix}
\end{figure}

Figure~\ref{fig:conf_matrix} provides the confusion matrix for the test set predictions, showing the distribution of correct and incorrect classifications. Notably, out of the total 1,500 test samples (750 fraud and 750 non-fraud), the model misclassifies only a small fraction (45 non-fraud labeled as fraud and 25 fraud labeled as non-fraud), aligning with the high F1 score.

\subsubsection{Discussion of Overfitting and Generalization}

One notable observation from our experiments is that the training accuracy is significantly higher than the validation and test accuracies, and similarly, the training loss is much lower than the validation loss. Despite applying standard regularization techniques such as dropout (with a rate of 0.3) and weight decay ($1 \times 10^{-4}$), this gap persists. This discrepancy suggests that the model may be overfitting to the training data.

Several factors might contribute to this behavior:
\begin{itemize}
    \item \textbf{Dataset Size and Diversity:} The relatively small dataset (10,000 samples with a balanced 50-50 distribution) might limit the model's ability to learn a generalizable representation. A larger and more diverse dataset could help in reducing overfitting.
    \item \textbf{Model Expressivity:} The high expressivity of the hybrid model, especially due to the quantum circuit's ability to capture complex feature interactions via entanglement and superposition \cite{Biamonte2017, Schuld2019}, may cause it to fit the training data very closely, capturing noise and idiosyncratic patterns that do not generalize well.
    
\end{itemize}

These challenges are not uncommon in hybrid quantum-classical models and warrant further investigation. Future work will focus on exploring additional regularization techniques, such as more advanced dropout schemes, data augmentation, or ensemble methods, as well as leveraging larger datasets to improve generalization.

\subsection{Classical LSTM Results for Benchmark}
\label{sec:classical_lstm_results}

To establish a baseline against our hybrid quantum-classical model, we also trained a purely classical LSTM 
on the same dataset. Unlike the quantum model, this LSTM directly connects its final hidden state to a fully 
connected layer for fraud classification.

\subsubsection{Training and Validation}
During training, the classical LSTM's accuracy improved from approximately 70\% to over 94\% in 80 epochs, 
while validation accuracy rose from 83\% to about 93\%. This rapid convergence indicates that the LSTM 
learned relevant patterns effectively. Notably, we observe no clear signs of overfitting: the training 
and validation metrics remain closely aligned throughout training.

\subsubsection{Loss and Accuracy Curves}
\begin{figure}[ht]
    \centering
    \begin{subfigure}[b]{0.48\linewidth}
        \centering
        \includegraphics[width=\linewidth]{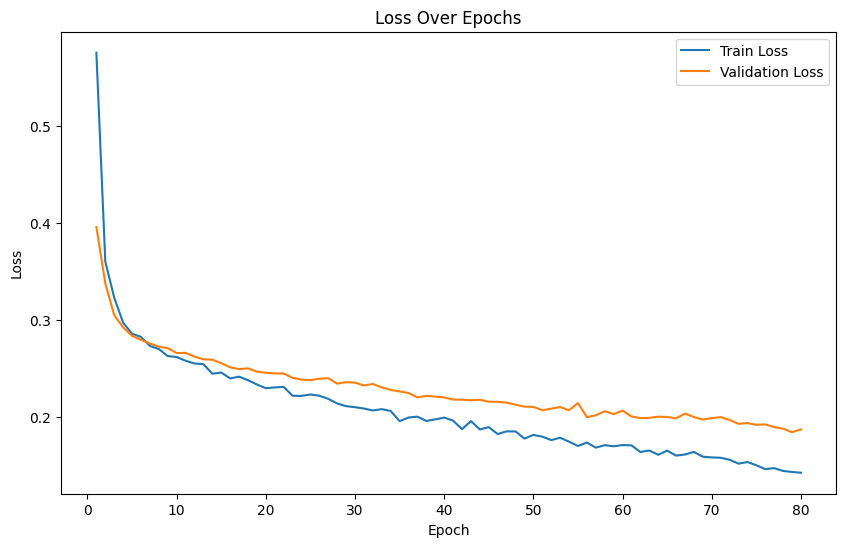}
        \caption{Loss Over Epochs}
    \end{subfigure}
    \hfill
    \begin{subfigure}[b]{0.48\linewidth}
        \centering
        \includegraphics[width=\linewidth]{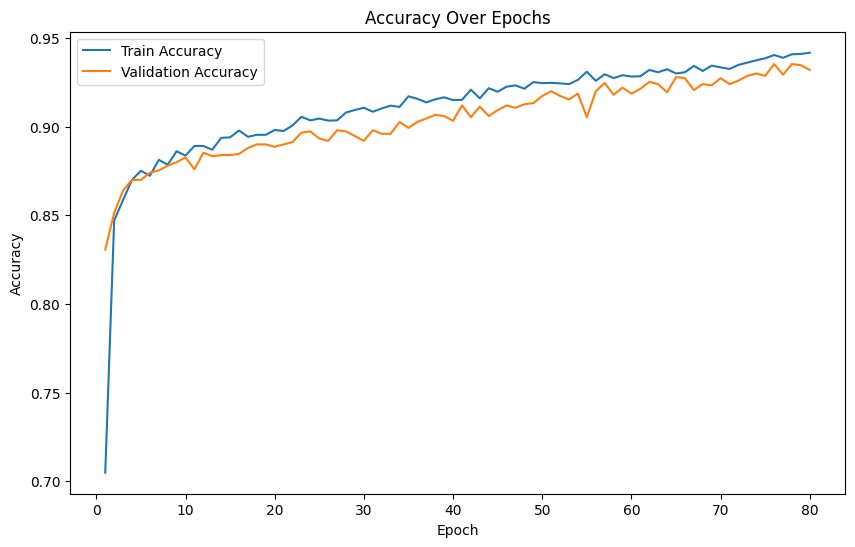}
        \caption{Accuracy Over Epochs}
    \end{subfigure}
    \caption{Training and validation loss and accuracy for the Classical LSTM. 
    The model converges smoothly, with no evident overfitting.}
    \label{fig:classical_lstm_curves}
\end{figure}

Figure~\ref{fig:classical_lstm_curves} displays the loss and accuracy curves over 80 epochs. The loss 
consistently decreases from 0.57 to 0.14, and accuracy steadily increases, reflecting a robust learning 
process. The close match between training and validation curves further confirms the model's generalization 
capability on this dataset.

\subsubsection{Time Per Epoch and Computational Analysis}
\begin{figure}[ht]
    \centering
    \begin{subfigure}[b]{0.48\linewidth}
         \centering
         \includegraphics[width=\linewidth]{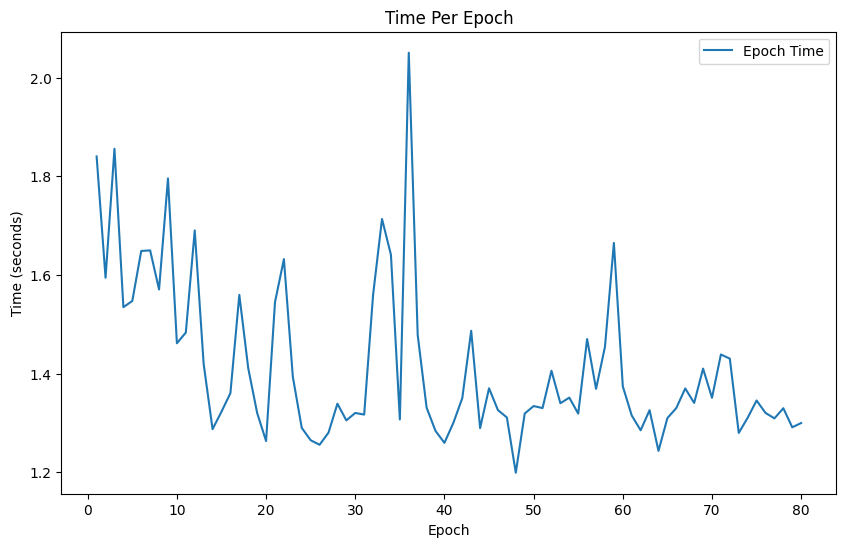} 
         \caption{Time Per Epoch}
         \label{fig:classical_time}
    \end{subfigure}
    \hfill
    \begin{subfigure}[b]{0.48\linewidth}
         \centering
         \includegraphics[width=\linewidth]{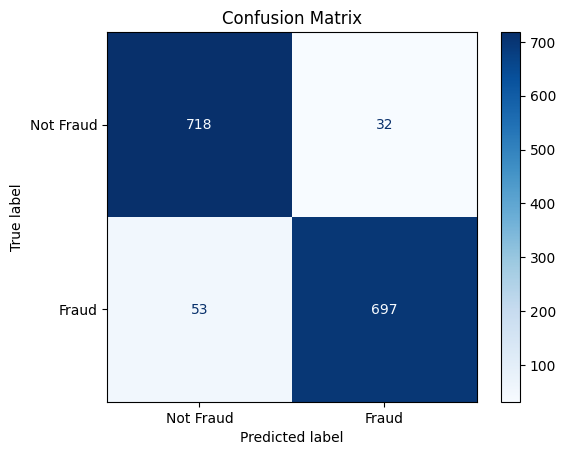} 
         \caption{Confusion Matrix}
         \label{fig:classical_conf_matrix}
    \end{subfigure}
    \caption{(a) Per-epoch training time for the Classical LSTM, with most epochs under 2 seconds, and (b) the corresponding confusion matrix on the test set.}
    \label{fig:combined_classical}
\end{figure}

On average, the classical LSTM required under 2 seconds per epoch to train, which is significantly faster than the hybrid quantum model. This efficiency is primarily due to the absence of quantum circuit simulation overhead, which typically incurs additional computational cost. Figure~\ref{fig:classical_time} illustrates the per-epoch training time over 80 epochs. The training times remain relatively stable with only minor fluctuations, likely due to variations in system load and batch processing times.

The computational efficiency of the classical LSTM allows for rapid experimentation and hyperparameter tuning. However, while it offers this speed advantage, purely classical models may capture fewer complex feature correlations compared to quantum-enhanced architectures.

In addition to the timing analysis, we also evaluated the model’s classification performance using a confusion matrix. Figure~\ref{fig:classical_conf_matrix} shows that the classical LSTM achieves robust performance with the majority of fraudulent and non-fraudulent transactions correctly classified. The relatively low number of misclassifications further confirms the model's effectiveness and balanced performance.

\subsubsection{Comparison of Quantum Hybrid and Classical LSTM Models}
To provide a comprehensive comparison between the quantum hybrid model and the classical LSTM model, Table~\ref{tab:comparison_results} summarizes key performance metrics along with computational efficiency measures, including the average time per epoch.

\begin{table}[ht]
    \centering
    \caption{Comparison of Quantum Hybrid and Classical LSTM Models on Fraud Detection}
    \label{tab:comparison_results}
    \begin{tabular}{lcc}
        \hline
        \textbf{Metric} & \textbf{Quantum Hybrid} & \textbf{Classical LSTM} \\
        \hline
        Accuracy (\%)                  & 95.33 & 94.33 \\
        Precision (\%)                 & 94.16 & 95.61 \\
        Recall (\%)                    & 96.67 & 92.93 \\
        F1 Score (\%)                  & 95.39 & 94.25 \\
        Evaluation Time (s)            & 6.32  & 0.07 \\
        Average Time per Epoch (s)     & 55.00 & 1.50 \\
        \hline
    \end{tabular}
\end{table}
\noindent To further evaluate the model's scalability and performance, we increased the dataset size and experimented with larger qubit configurations. Specifically, we trained the model with \textbf{10 qubits} on \textbf{35k samples} and \textbf{12 qubits} on \textbf{30k samples}, comparing the results to the original \textbf{10 qubits with 10k samples} configuration. As expected, the training time increased significantly with the larger datasets and higher qubit count. For \textbf{10 qubits with 35k samples}, the model required approximately \textbf{165--180 seconds per epoch} (with occasional spikes), while the \textbf{12-qubit configuration with 30k samples} took around \textbf{4--5 minutes per epoch}. 

In terms of performance, the larger datasets led to improved metrics, with the \textbf{accuracy}, \textbf{recall}, and \textbf{F1 score} all showing noticeable gains. The \textbf{10-qubit model with 35k samples} achieved an accuracy of \textbf{98.38\%}, recall of \textbf{99.81\%}, and an F1 score of \textbf{98.40\%}, while the \textbf{12-qubit model with 30k samples} achieved an accuracy of \textbf{98.04\%}, recall of \textbf{100.00\%}, and an F1 score of \textbf{98.08\%}. These results demonstrate that \textbf{increasing the number of samples} effectively reduces the overfitting gap, leading to \textbf{better generalization}. However, it is also evident that while larger qubit models yield marginally improved metrics, they incur a \textbf{significant computational cost}, making the trade-off between performance and efficiency an important consideration for real-world deployment.

\begin{table}[ht]
    \centering
    \caption{Performance and Training Time Comparison for Different Qubit and Dataset Configurations}
    \label{tab:qubit_comparison}
    \begin{tabular}{lccc}
        \hline
        \textbf{Configuration}         & \textbf{Accuracy (\%)}   & \textbf{Recall (\%)}   & \textbf{F1 Score (\%)}       \\
        \hline
        10 qubits, 10k samples         & 95.33                    & 96.67                   & 95.39                        \\
        10 qubits, 35k samples         & 98.38                    & 99.81                   & 98.40                        \\
        12 qubits, 30k samples         & 98.04                    & 100.00                  & 98.08                        \\
        \hline
        \multicolumn{4}{c}{\textbf{Average Time per Epoch}} \\
        \hline
        10 qubits, 10k samples         & \multicolumn{3}{c}{55 sec}         \\
        10 qubits, 35k samples         & \multicolumn{3}{c}{165--180 sec}   \\
        12 qubits, 30k samples         & \multicolumn{3}{c}{4--5 min}       \\
        \hline
    \end{tabular}
\end{table}

\section{Conclusion}
\label{sec:conclusion}

In this study, we presented a novel hybrid quantum-classical neural network for fraud detection that integrates classical LSTM layers with a variational quantum circuit. Our experimental results demonstrate that the quantum hybrid model achieves a very high training accuracy of 99.69\% and a test accuracy of 95.33\%, indicating its ability to effectively capture complex temporal and feature-level dependencies in transaction data. While the classical LSTM model offers faster per-epoch training and inference, our hybrid approach outperforms it in key predictive metrics, particularly in recall and F1 score, which are crucial for fraud detection applications.

Our work is novel not only in terms of its competitive computational efficiency, averaging 45--65 seconds per epoch, but also in its innovative architecture. The model leverages quantum encoding through \emph{AngleEmbedding} and strongly entangling layers, enabling it to explore high-dimensional feature spaces via quantum phenomena such as superposition and entanglement. This design enhances the expressivity of the model beyond what is possible with classical architectures alone. Despite some degree of overfitting indicated by the gap between training and test accuracies, our results highlight significant performance improvements compared to existing studies, where hybrid quantum models typically require several minutes per epoch or demonstrate lower overall accuracy .

Future work will focus on addressing the remaining overfitting issues by exploring additional regularization techniques, further optimizing quantum circuit parameters, and expanding the dataset to improve generalization, and on validating the model on actual quantum hardware. In particular, testing with a higher number of qubits is anticipated to enhance the expressivity of the quantum circuit; however, such scaling on real quantum devices will require advanced error mitigation strategies (such as zero-noise extrapolation or probabilistic error cancellation) to manage hardware-induced noise and decoherence. In summary, our proposed hybrid architecture offers a promising avenue for practical fraud detection by combining high predictive performance with enhanced feature representation and improved computational efficiency.



\begin{thebibliography}{10}
\providecommand{\url}[1]{#1}
\csname url@samestyle\endcsname
\providecommand{\newblock}{\relax}
\providecommand{\bibinfo}[2]{#2}
\providecommand{\BIBentrySTDinterwordspacing}{\spaceskip=0pt\relax}
\providecommand{\BIBentryALTinterwordstretchfactor}{4}
\providecommand{\BIBentryALTinterwordspacing}{\spaceskip=\fontdimen2\font plus
\BIBentryALTinterwordstretchfactor\fontdimen3\font minus \fontdimen4\font\relax}
\providecommand{\BIBforeignlanguage}[2]{{%
\expandafter\ifx\csname l@#1\endcsname\relax
\typeout{** WARNING: IEEEtran.bst: No hyphenation pattern has been}%
\typeout{** loaded for the language `#1'. Using the pattern for}%
\typeout{** the default language instead.}%
\else
\language=\csname l@#1\endcsname
\fi
#2}}
\providecommand{\BIBdecl}{\relax}
\BIBdecl

\bibitem{heryadi2017learning}
Y.~Heryadi and H.~L. H.~S. Warnars, ``Learning temporal representation of transaction amount for fraudulent transaction recognition using cnn, stacked lstm, and cnn-lstm,'' in \emph{2017 IEEE International Conference on Cybernetics and Computational Intelligence (CyberneticsCom)}.\hskip 1em plus 0.5em minus 0.4em\relax IEEE, 2017, pp. 84--89.

\bibitem{guo2018learning}
J.~Guo, G.~Liu, Y.~Zuo, and J.~Wu, ``Learning sequential behavior representations for fraud detection,'' in \emph{2018 IEEE international conference on data mining (ICDM)}.\hskip 1em plus 0.5em minus 0.4em\relax IEEE, 2018, pp. 127--136.

\bibitem{cheng2020spatio}
D.~Cheng, S.~Xiang, C.~Shang, Y.~Zhang, F.~Yang, and L.~Zhang, ``Spatio-temporal attention-based neural network for credit card fraud detection,'' in \emph{Proceedings of the AAAI conference on artificial intelligence}, vol.~34, no.~01, 2020, pp. 362--369.

\bibitem{alghofaili2020financial}
Y.~Alghofaili, A.~Albattah, and M.~A. Rassam, ``A financial fraud detection model based on lstm deep learning technique,'' \emph{Journal of Applied Security Research}, vol.~15, no.~4, pp. 498--516, 2020.

\bibitem{benchaji2021enhanced}
I.~Benchaji, S.~Douzi, B.~El~Ouahidi, and J.~Jaafari, ``Enhanced credit card fraud detection based on attention mechanism and lstm deep model,'' \emph{Journal of Big Data}, vol.~8, pp. 1--21, 2021.

\bibitem{grossi2022mixed}
M.~Grossi, N.~Ibrahim, V.~Radescu, R.~Loredo, K.~Voigt, C.~Von~Altrock, and A.~Rudnik, ``Mixed quantum--classical method for fraud detection with quantum feature selection,'' \emph{IEEE Transactions on Quantum Engineering}, vol.~3, pp. 1--12, 2022.

\bibitem{kyriienko2022unsupervised}
O.~Kyriienko and E.~B. Magnusson, ``Unsupervised quantum machine learning for fraud detection,'' \emph{arXiv preprint arXiv:2208.01203}, 2022.

\bibitem{innan2024financial}
N.~Innan, A.~Sawaika, A.~Dhor, S.~Dutta, S.~Thota, H.~Gokal, N.~Patel, M.~A.-Z. Khan, I.~Theodonis, and M.~Bennai, ``Financial fraud detection using quantum graph neural networks,'' \emph{Quantum Machine Intelligence}, vol.~6, no.~1, p.~7, 2024.

\bibitem{wang2022integrating}
H.~Wang, W.~Wang, Y.~Liu, and B.~Alidaee, ``Integrating machine learning algorithms with quantum annealing solvers for online fraud detection,'' \emph{Ieee Access}, vol.~10, pp. 75\,908--75\,917, 2022.

\bibitem{innan2024qfnn}
N.~Innan, A.~Marchisio, M.~Bennai, and M.~Shafique, ``Qfnn-ffd: Quantum federated neural network for financial fraud detection,'' \emph{arXiv preprint arXiv:2404.02595}, 2024.

\bibitem{khan2024quantum}
S.~Z. Khan, N.~Muzammil, S.~Ghafoor, H.~Khan, S.~M.~H. Zaidi, A.~J. Aljohani, and I.~Aziz, ``Quantum long short-term memory (qlstm) vs. classical lstm in time series forecasting: a comparative study in solar power forecasting,'' \emph{Frontiers in Physics}, vol.~12, p. 1439180, 2024.

\bibitem{Sim_2019}
\BIBentryALTinterwordspacing
S.~Sim, P.~D. Johnson, and A.~Aspuru‐Guzik, ``Expressibility and entangling capability of parameterized quantum circuits for hybrid quantum‐classical algorithms,'' \emph{Advanced Quantum Technologies}, vol.~2, no.~12, Oct. 2019. [Online]. Available: \url{http://dx.doi.org/10.1002/qute.201900070}
\BIBentrySTDinterwordspacing

\bibitem{Hochreiter1997}
S.~Hochreiter and J.~Schmidhuber, ``Long short-term memory,'' \emph{Neural Computation}, vol.~9, no.~8, pp. 1735--1780, 1997.

\bibitem{Goodfellow2016}
I.~Goodfellow, Y.~Bengio, and A.~Courville, \emph{Deep Learning}.\hskip 1em plus 0.5em minus 0.4em\relax MIT Press, 2016, \url{http://www.deeplearningbook.org}.

\bibitem{Chen2022LSTM}
S.-Y. Chen, S.~Yoo, Y.-L. Fang \emph{et~al.}, ``Quantum long short-term memory,'' \emph{ICASSP 2022 - 2022 IEEE International Conference on Acoustics, Speech and Signal Processing (ICASSP)}, pp. 8622--8626, 2022.

\bibitem{Biamonte2017}
J.~Biamonte, P.~Wittek, N.~Pancotti, P.~Rebentrost, N.~Wiebe, and S.~Lloyd, ``Quantum machine learning,'' \emph{Nature}, vol. 549, no. 7671, pp. 195--202, 2017.

\bibitem{Schuld2019}
M.~Schuld, N.~Killoran, V.~Bergholm, J.~Izaac, and C.~Weedbrook, ``Quantum machine learning in feature hilbert spaces,'' \emph{Physical Review A}, vol.~99, no.~3, p. 032331, 2019.

\bibitem{Havlicek2019}
V.~Havl{\'\i}{\v{c}}ek, A.~C{\'o}rcoles, K.~Temme, D.~Harburn, A.~Kandala, J.~Chow, and J.~Gambetta, ``Supervised learning with quantum-enhanced feature spaces,'' \emph{Nature}, vol. 567, no. 7747, pp. 209--212, 2019.

\bibitem{Mitarai2018}
K.~Mitarai, M.~Negoro, M.~Kitagawa, and K.~Fujii, ``Quantum circuit learning,'' \emph{Physical Review A}, vol.~98, no.~3, p. 032309, 2018.

\bibitem{Bergholm2018}
V.~Bergholm, J.~Izaac, M.~Schuld, C.~Gogolin, S.~Ahmed, V.~Ajith \emph{et~al.}, ``{PennyLane}: Automatic differentiation of hybrid quantum-classical computations,'' 2018, \url{https://pennylane.ai}.

\bibitem{Chen2022Quantum}
S.-Y. Chen, Y.-L. Fang, and S.~Yoo, ``Quantum long short-term memory,'' \emph{arXiv preprint arXiv:2202.07431}, 2022.

\bibitem{Kingma2014}
D.~P. Kingma and J.~Ba, ``Adam: A method for stochastic optimization,'' arXiv preprint arXiv:1412.6980, 2014.

\end{thebibliography}
\end{document}